\documentclass[%
 reprint,
superscriptaddress,
 amsmath,amssymb,
 aps,
 pra    ,
]{revtex4-2}

\usepackage{graphicx}
\usepackage{dcolumn}
\usepackage{bm}
\usepackage{xcolor}
\usepackage{bbm}
\usepackage{physics}
\usepackage{appendix}
\usepackage{soul}
\usepackage{epsfig}
\usepackage{float}
\usepackage{multirow}
\usepackage{braket}
\usepackage{lipsum}
\usepackage{hyperref}
\hypersetup{
    colorlinks=true,
    linkcolor=blue,
    filecolor=magenta,      
    urlcolor=blue,
    citecolor = blue,
}

\begin{document}
\preprint{APS/123-QED}

\title{Equivalence regimes for geometric quantum discord and local quantum uncertainty} 

\author{Oscar Cordero}
\affiliation{ICFO—Institut de Ciencies Fotoniques, the Barcelona Institute of Science and Technology, 08860 Castelldefels (Barcelona), Spain}

\author{Arturo Villegas}
\affiliation{ICFO—Institut de Ciencies Fotoniques, the Barcelona Institute of Science and Technology, 08860 Castelldefels (Barcelona), Spain}

\author{Juan-Rafael Alvarez}
\affiliation{Clarendon Laboratory, University of Oxford, Parks Road, Oxford OX1 3PU, United Kingdom}

\author{Roberto de J. Le\'on-Montiel}
\affiliation{Instituto de Ciencias Nucleares, Universidad Nacional Autónoma de M\'exico, Apartado Postal 70-543, 04510 Cd. Mx., M\'exico}

\author{M. H. M. Passos}
\affiliation{ICFO—Institut de Ciencies Fotoniques, the Barcelona Institute of Science and Technology, 08860 Castelldefels (Barcelona), Spain}
\email{marcello.passos@icfo.eu}

\author{Juan P. Torres}
\affiliation{ICFO—Institut de Ciencies Fotoniques, the Barcelona Institute of Science and Technology, 08860 Castelldefels (Barcelona), Spain}
\affiliation{Department of Signal Theory and Communications, Universitat Politecnica de Catalunya, 08034 Barcelona, Spain}

\date{\today}

\begin{abstract}
The concept of quantum discord aims at unveiling quantum correlations that go beyond those described by entanglement. Its  original formulation [J. Phys. A {\bf 34}, 6899 (2001); Phys. Rev. Lett {\bf 88}, 017901 (2002)] is difficult to compute even for the simplest case of two-qubits systems. Alternative formulations have been developed to address this drawback, such as the geometric measure of quantum discord [Phys. Rev. A {\bf 87}, 062303 (2013)] and the local quantum uncertainty [Phys. Rev. Lett {\bf 110}, 240402 (2013)] that can be evaluated in closed form for some quantum systems, such as two-qubit systems. We show here that these two measures of quantum discord are equivalent for $2 \times D$ dimensional bipartite quantum systems. By considering the relevant example of N00N states for phase estimation in lossy environments, we also show that both metrics of quantum discord quantify the decrease of quantum Fisher information of the phase estimation protocol. Given their ease of computation in $2 \times D$ bipartite systems, the geometric measure of quantum discord and the local quantum uncertainty demonstrate their relevance as computable measures of quantum discord.
\end{abstract}

\keywords{Non-classical correlations, coherence, quantum discord, entanglement}
\maketitle

\section{\label{sec:Introduction}Introduction}
The quantum correlations embedded in entangled states are a resource that facilitates the design of new protocols for parameter estimation. Relative to coherent states, usually considered as benchmark states, entangled states can show enhanced resolution. One paradigmatic example of such states  used for quantum-enhanced sensing are N00N states, which allow the estimation of an unknown phase with a resolution that scales as $1/N$, where $N$ is the average number of photons. This is an improvement with respect to the scaling provided by coherent states, that goes as $\sim 1/\sqrt{N}$.

Quantum correlations that go beyond those described by entanglement, e.g., quantum correlations in separable states, can also offer a quantum advantage by enhancing the resolution for estimating unknown parameters in a quantum system \cite{vedral2011}. Henderson and Vedral \cite{henderson2001}, and Ollivier and Zurek \cite{ollivier2002} introduced the concept of quantum discord to quantify those correlations. They noticed that while there are two equivalent expressions for the mutual information of two random variables that give the same result, their generalizations for measuring the correlations between two quantum systems may yield different results.

The original formulation of quantum discord is difficult to compute \cite{huang2014} even for the important but simplest case of two-qubit systems \cite{luo2008,ali2010,chen2011}. This has led to alternative formulations of the concept that still fulfill a set of conditions expected for a good measure of quantum correlations \cite{modi2012} while being more easily computable in certain scenarios of interest.

One of these alternatives is the geometric measure of quantum discord, or geometric quantum discord (GQD) for short. It is based on the assumption that a bipartite quantum state $\rho^{A B}$ has zero discord \cite{introgeometricdiscord,datta2011,acin2009,ferreira2018} if and only if there is a von Neumann measurement $\left\{\Pi_{k}^{A}\right\}=\left|u_{k}\right\rangle\left\langle u_{k}\right|$ on the subspace $A$ such that $\sum_{k}\left(\Pi_{k}^{A} \otimes I^{B}\right) \rho\left(\Pi_{k}^{A} \otimes I^{B}\right)=\rho .$ Here $I^{B}$ designates
the identity operator in the subspace $B$. \textcolor{black}{We restrict ourselves to von Neumann measurements \cite{Busch2009,luoandsun2017}, so all projectors $\Pi_{k}^{A}$ are one-dimensional. In this case we can write the projectors $\Pi_{k}^{A}$ in terms of a set of vectors $\left\{\left|u_{k}\right\rangle\right\}$ that is a basis in subspace $A$}. 

This implies that zero-discord quantum states are of the form $\rho=\sum_{k} p_{k}\left|u_{k}\right\rangle\left\langle u_{k}\right| \otimes \rho_{k}^{B}$, where $\rho_{k}^{B}$ are density matrices in subspace $B$ and $p_{k}$ are positive real numbers with $\sum_{k} p_{k}=1 .$ These states are sometimes termed as \textit{classical-quantum} \cite{adesso2016}. From the definition of \textit{classical-quantum states}, it naturally follows that the geometric quantum discord is the minimum distance (square norm in the Hilbert-Schmidt space) between the quantum state $\rho$ and the closest \textit{classical-quantum state} $\sum_{k}\left(\Pi_{k}^{A} \otimes I^{B}\right) \rho\left(\Pi_{k}^{A} \otimes I^{B}\right)$.

Such a definition for GQD might show some drawbacks \cite{piani2012} since it can increase under local operations of the party $B$ that is not measured. This undesirable effect can be corrected \cite{ChangLuo2013} if one substitutes the density matrix $\rho$ by $\rho^{1/2}$, so that the GQD is now the minimum distance (square norm in the Hilbert-Schmidt space) between $\rho^{1/2}$ and $\sum_k \left( \Pi_k^A \otimes I^B \right)\, \rho^{1/2}\, \left( \Pi_k^A \otimes I^B \right)$. This is the version of geometric quantum discord that we use throughout this paper. One major advantage of this expression is that it  can be calculated in closed form for  quantum bipartite systems of dimension $2 \times D$  \cite{ChangLuo2013,LuoFu2012}. 

Interestingly, the very same year that the previous correction of the geometric discord was reported,  Girolami, Tufarelli and Adesso \cite{Girolami2013} introduced the local quantum uncertainty (LQU), a new formulation of quantum discord defined as follows: given a specific von Neumann measurement where each projector $\Pi_k^A$ is assigned an eigenvalue $\lambda_k$ (all $\lambda_k$ are different), the LQU is the minimum over all possible ensembles $\left\{ \Pi_k^A\right\}$ of the Wigner-Yanase Skew information, $I$ \cite{Wigner910}:
\begin{equation}
    I=-\frac{1}{2} \text{Tr} \left\{ \left[ \rho^{1/2}, M\right]^2 \right\}.
\end{equation}
Here $M=\left( \sum_k \lambda_k \Pi_k^A\right)\,\otimes  I^B$ and $I^B$ is the identity on subspace $B$. Again, as in the case of the geometric quantum discord discussed above, one important advantage of LQU is that it can be calculated in closed form for $2 \times D$ quantum bipartite systems.

For a given von Neumann measurement $\left\{ \Pi_k^A \otimes I^B \right\}$, one can define its quantum uncertainty as $Q=\sum_k I_k$, where
\begin{equation}
    I_k=-\frac{1}{2} \text{Tr} \left\{ \left[ \rho^{1/2}, \Pi_k^A \otimes I^B\right]^2 \right\}.
\end{equation}
It turns out that the GQD is the \textcolor{black}{minimum} of the quantum uncertainty $Q$ over all possible von Neumann measurements. This introduces a revealing link between the LQU and the GQD formulations of the quantum discord through the use of similar expressions of the Wigner-Yanase Skew information \cite{luoandsun2017}. \textcolor{black}{In a given von Neumann measurement, characterized by a set of one-dimensional operators $\left\{ \Pi_k^A \right\}$, each one associated with a possible experimental outcome, the intrinsic statistical error associated with the measurement has a quantum contribution. The Skew information, a measure of the non-commutativity between the quantum state  $\rho$ and the set $\left\{ \Pi_k^A \otimes I^B \right\}$, can be used to quantify this quantum uncertainty.  In this context, the local quantum uncertainty and the geometric discord can be understood as the minimum quantum uncertainty that one can have among all possible von Neumann measurements. However, they differ in how they evaluate the quantum uncertainty. The geometric discord considers the sum of the quantum uncertainties associated with each outcome $\Pi_k^A \otimes I^B$, while the local quantum uncertainty considers the quantum uncertainty associated to an operator that describes the global measurement, $M=\left( \sum_k \lambda_k  \Pi_k^A \right) \otimes I^B$, where $\lambda_k$ are eigenvalues associated with each possible outcome of the measurement.}

\textcolor{black}{The two quantum discord metrics considered above, namely the local quantum uncertainty and the geometric quantum discord, fulfil similar requirements that the original discord definition does, which make them good discord metrics \cite{modi2012,Girolami2013}. These discord quantifiers are non-negative, invariant under local unitary transformations, they yield zero only for quantum-classical states and the discord reduces to an entanglement monotone, characterized by the marginal entropy of subsystem $A$, for pure states.}

\textcolor{black}{As the geometric discord and the local quantum uncertainty can be both explained as the minimum quantum uncertainty that can be attained in a von Neumann measurement, one might wonder whether they are the same discord metric, at least for certain scenarios}. In this paper, we demonstrate that for bipartite quantum systems whose dimensionality is $2\times D$, the two aforementioned metrics of quantum discord are indeed the same, although this may not be true for systems with other dimensions. Moreover, we take advantage of the fact that both measures can be evaluated in closed form, in sharp contrast to other alternative formulations of quantum discord \cite{vedral2011}.

Finally, we show an example of the potential usefulness of GQD and LQU by evaluating the quantum Fisher information of N00N states for phase estimation in a lossy environment. \textcolor{black}{The use of quantum systems in sensing and imaging applications provides a unique tool to develop new parameter estimation schemes with enhanced resolution. However, quantum systems experiencing losses are fragile. This can lead to a worsening of the resolution achievable, thus reducing the quantum advantage observed for the lossless case. We can use several measures to characterize the effect of losses, i. e., negativity and quantum discord, but it is not clear in principle which is the most convenient or informative in each scenario.}

For one-parameter estimation, the Cram\'er-Rao bound given by the quantum Fisher information \cite{Helstrom1969} is attainable, so it is a good measure of the resolution enhancement provided by a protocol making use of a specific quantum state \cite{fujiwara2005,matsumoto2005}. Remarkably, we demonstrate that the decrease of quantum Fisher information under the presence of losses, with respect to the ideal case with no losses, is precisely the geometric quantum discord. \textcolor{black}{In this sense, the quantum discord is more informative than negativity concerning the spatial resolution achievable under the present of loss, as given by the quantum Fisher information}.

\section{Equivalence between LQU and GQD for $2 \times D$ systems} 
The quantum uncertainty $Q$ defined in \cite{luoandsun2017}, whose minimum yields the GQD, can be written as $Q=\sum_j I_j$ where 
\begin{eqnarray}
& & I_j=-\frac{1}{2}\,\text{Tr} \left\{ \left[ \rho^{1/2},\Pi_j^A \otimes I^B \right]^2 \right\} \nonumber \\
& & =\text{Tr} \left[ \rho \left( \Pi_j^A \right)^2 \right]- \text{Tr}  \left( \rho^{1/2} \Pi_j^A \, \rho^{1/2} \Pi_j^A \right)  =\text{Tr}_B\, V_j, 
\end{eqnarray}
and $V_j$ is defined as
\begin{equation}
V_j=\langle u_j |\rho|u_j \rangle- \langle u_j |\rho^{1/2}|u_j\rangle \langle u_j|\rho^{1/2}|u_j \rangle. 
\end{equation}
If we make use of the resolution of the identity on subspace $A$, i.e., $\sum_i |u_i\rangle \langle u_i|=I^A$, we obtain that
\begin{equation}
    Q=\sum_j \text{Tr}_B\, V_j= 2\sum_{j<k} \text{Tr}_B\, V_{jk}, \label{Q}
\end{equation}
where
\begin{equation}
V_{jk}=\langle u_j |\rho^{1/2}|u_k\rangle \langle u_k|\rho^{1/2}|u_j \rangle, 
\end{equation}
and $V_{jk}=V_{kj}$.

\begin{figure*}[t!]
\includegraphics[width=\linewidth]{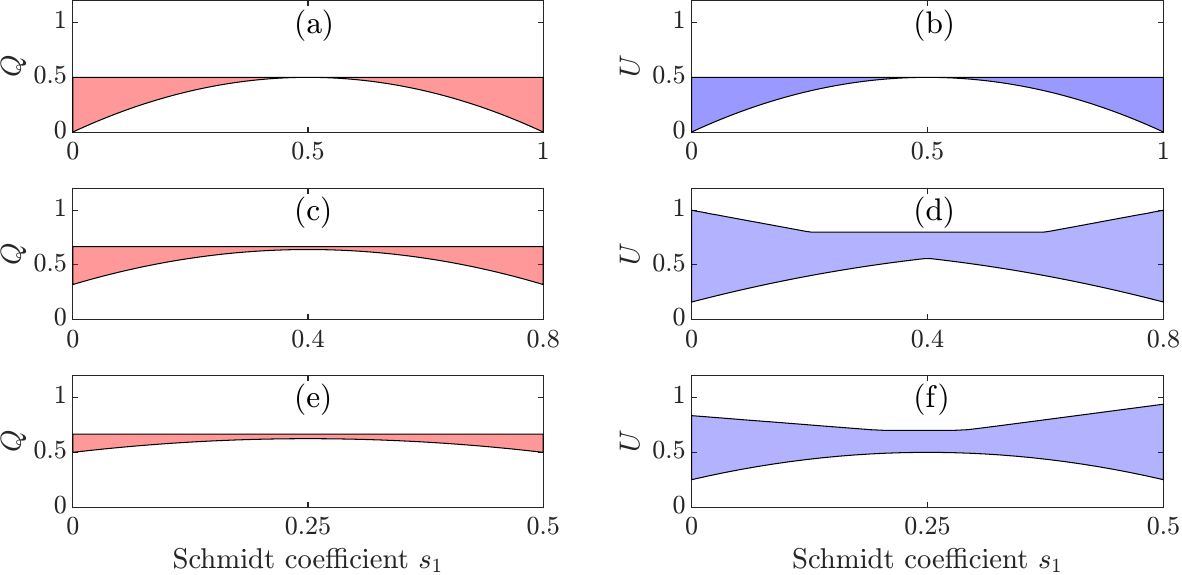}             
\caption{\label{figure1} All possible values of the Quantum uncertainties $Q$ (left) and $U$ (right) as a function of the Schmidt coefficients $s_1$ for fixed values of $s_2$, as indicated in the plot. (a) and (b) show results for $2 \times D$  systems, while (c)-(f) show results for $3 \times D$ systems with $\lambda_1=4, \lambda_2=3$ and $\lambda_3=2$. (c) and (d): Schmidt coefficient $s_2=0.2$; (e) and (f): Schmidt coefficient $s_2=0.5$. \textcolor{black}{The Schmidt coefficients are dimensionless.}}
\end{figure*}

In a similar vein, the quantum uncertainty $U$ defined in \cite{Girolami2013}, whose minimum yields the LQU, can be written as
\begin{eqnarray}
& & U=\text{Tr}_B \left\{  \sum_j \lambda_j^2 \langle u_j |\rho|u_j \rangle \right. \nonumber \\
& & \left. - \sum_{j,k}\lambda_j \lambda_k \langle u_j |\rho^{1/2}|u_k\rangle \langle u_k|\rho^{1/2}|u_j \rangle \right\} \nonumber \\
& &  = \sum_j \lambda_j^2 \text{Tr}_B\,V_j-2 \sum_{jk} \text{Tr}_B\,\lambda_j \lambda_k V_{jk} \nonumber \\
& & = \sum_{j< k} (\lambda_j^2+\lambda_k^2)\, \text{Tr}_B V_{jk} -2\sum_{j<k} \lambda_j \lambda_k \text{Tr}_B\,V_{jk} \nonumber \\
& & =\sum_{j<k} (\lambda_j-\lambda_k)^2 \text{Tr}_B\, V_{jk}, \label{U}
\end{eqnarray}
where $\lambda_j$ corresponds to the eigenvalue of the $j$-th projector constituting a von Neumann measurement.

Equations (\ref{Q}) and (\ref{U}) are valid for arbitrary dimensions of the Hilbert spaces of the bipartite quantum states, and for any quantum state described by density matrix $\rho$. For a Hilbert space with dimension $2 \times D$ the key observation is that 
\begin{eqnarray}
& & \langle u_1 |\rho|u_1 \rangle- \langle u_1 |\rho^{1/2}|u_1\rangle \langle u_1|\rho^{1/2}|u_1 \rangle \nonumber \\
& & = \langle u_1 |\rho^{1/2}|u_2\rangle \langle u_2|\rho^{1/2}|u_1 \rangle \nonumber \\
& & = \langle u_2 |\rho|u_2 \rangle- \langle u_2 |\rho^{1/2}|u_2\rangle \langle u_2|\rho^{1/2}|u_2 \rangle.
\end{eqnarray}
so that $V_1=V_2=V_{12}$. In this case,
\begin{equation}
U=(\lambda_1-\lambda_2)^2 \text{Tr}_B\, V_{12}=\frac{(\lambda_1-\lambda_2)^2}{2} Q.\label{proportionalQU}
\end{equation}
Equation (\ref{proportionalQU}) shows that the quantum uncertainties $Q$ and $U$ are proportional to each other, thus implying that the measures of quantum discord that derive from them are indeed equivalent for bipartite systems of dimension $2 \times D$. 

\section{Non-equivalence between LQU and GQD in systems with arbitrary dimensions}
\textcolor{black}{In this section we want to demonstrate that in bipartite systems where the dimension of both subsystems is greater than $2$, the LQU and GQD are not proportional to each other. For the sake of simplicity, we restrict ourselves to comparing the values of $Q$ and $U$ for pure states in Hilbert spaces of dimensions $2\times D$ and $3\times D$}. 

We start by noticing that any pure bipartite quantum state can be written as a Schmidt decomposition
\begin{equation}
|\Psi\rangle=\sum_m \sqrt{s_m} |\alpha_m \rangle |\beta_m \rangle,
\end{equation}
where $\left\{ \alpha_m \right\}$ is a basis in subspace $A$, $\left\{ \beta_m \right\}$ is a basis in subspace $B$ and $\left\{ s_j \right\}$ are the Schmidt coefficients, with the normalization condition $\sum_j s_j=1$. We can easily derive that
\begin{equation}
    \text{Tr}_B\,V_{jk}=\big[ \sum_m s_m \big|\langle \alpha_m|u_j\rangle \big|^2 \big] \times \big[ \sum_n s_n \big|\langle \alpha_n|u_k\rangle \big|^2 \big].
\end{equation}
In Ref. \cite{ChangLuo2013} it was demonstrated that for pure states the von Neumann measurement that minimizes the quantum uncertainty $Q$ corresponds to choosing $|u_i \rangle \equiv |\alpha_i \rangle$. In this case $\text{Tr}_B\, V_{jk}=s_j s_k$ so the geometric quantum discord \textcolor{black}{for pure states} is $D_G=2\sum_{j<k} s_j s_k$. By making use of the normalization of the quantum state we obtain that $2\sum_{i<j} s_i s_j=1-\sum_i s_i^2$ so the quantum discord for pure states can also be written as $D_G=1-\sum_i s_i^2$, as reported in \cite{ChangLuo2013}.

The expression of the quantum uncertainty $U$ for pure states is
\begin{eqnarray}
& & U=\sum_{j<k} (\lambda_j-\lambda_k)^2 \big[ \sum_m s_m \big|\langle \alpha_m|u_j\rangle \big|^2 \big] \nonumber \\
& & \times \big[ \sum_n s_n \big|\langle \alpha_n|u_k\rangle \big|^2 \big].
\end{eqnarray}
We have performed extensive numerical simulations choosing many random von Neumann bases $\left\{ |u_i \rangle \right\}$ to calculate the range of possible values of the quantum uncertainties $Q$ and $U$. The von Neumann bases are obtained by choosing random unitary transformations $U$ of the bases $\left\{ |\alpha_i \rangle \right\}$ so that $\left\{ |u_i \rangle \right\}=U \left\{ |\alpha_i \rangle \right\}$. For $2 \times D$ and $3 \times D$ quantum systems, one can choose the most general unitary transformation as given in \cite{Rasin1997}.

Figure 1(a) shows all possible values of the quantum uncertainty $Q$ obtained numerically for a $2 \times D$ quantum system.  The solid lines correspond to the minimum value of $Q$, that is $D_G=2s_1(1-s_1)$, and the maximum value, $D_G=1-1/2=0.5$  \cite{ChangLuo2013}. Fig. 1(b) shows all possible values of $U$ for a $2 \times D$ quantum system with $(\lambda_1-\lambda_2)^2/2=1$. \textcolor{black}{As expected from the results obtained in Section II, Figs. 1(a) and (b) show the same results.} 

\textcolor{black}{Figs. 1(c) to 1(f) correspond to a $3 \times D$ system. The numerical simulations hereby presented show that the minimum of $U$ is attained for von Neumann measurements where the three orthogonal measurement projectors $\Pi_i^A$ ($i=1,2,3$) can be written as $\Pi_i^A=|\alpha_{p(i)} \rangle \langle \alpha_{p(i)}|$, where $p(i)$ designates the permutation $\left\{ 1,2,3 \right\} \longrightarrow \left\{ p(1), p(2), p(3) \right\}$ that yields the minimum value of $U$. We have six possibilities corresponding to the six different ways we can associate one vector of the set $|u_i \rangle$ with one vector of the set $|\alpha_i \rangle$.} The local quantum uncertainty is
\begin{equation}
LQU=\sum_{j<k} (\lambda_j-\lambda_k)^2 s_{p(j)} s_{p(k)}.
\end{equation}
The eigenvalue $\lambda_i$ that we associate to each von Neumann state $|\alpha_i \rangle$ now matters. \textcolor{black}{This is in contrast to the case of $Q$, where there is no eigenvalues associated to each outcome of a measurement and so all outcomes have the same weight.}

\textcolor{black}{Note that the maximum value of $Q$ for pure states is independent of the Schmidt coefficients $s_i$, and it is $1/2$ for $2 \times D$ systems and $2/3$ for $3 \times D$. On the other hand, Figs. 1(d) and (f) show that the maximum value of $U$ for $3 \times D$ systems may change for different values of the Schmidt coefficients- As a conclusion, such value does not depend only on the dimensions of the subsystems, which is the case of the quantum uncertainty $Q$.}

\textcolor{black}{Figure 2 shows how, for two specific set of values of the eigenvalues $\lambda_i$, the correspondence between vectors $|u_i \rangle$ and $|\alpha_i \rangle$ that give the minimum of quantum uncertainty $U$ varies for different values of $s_1$ and $s_2$. Each color in the figures stands for a different value of the minimum of $U$. Fig. 2(a) shows that for the case with eigenvalues $\lambda_1=2$, $\lambda_2=4$ and $\lambda_3=1$, when comparing the minimum of $U$ obtained for each value of $s_1$ and $s_2$, up to six different results are obtained. These six minimum values of $U$ can be obtained making use of the six possible permutations in Eq. (13). In Fig. 2(b) we consider the case with eigenvalues $\lambda_1=4$, $\lambda_2 = 3$ and $\lambda_3 = 2$. Now one can obtain up to three different minima of $U$ when considering all possible Schmidt coefficients.} 

\section{Geometric quantum discord of N00N states under non-symmetric losses }
To demonstrate the usefulness of the equivalence between GQD and LQU, we consider the relevant case of N00N states for phase estimation,
\begin{equation}
    \ket{\Psi}_{AB}=\frac{1}{\sqrt{2}}\Big(\ket{N}_A\ket{0}_B+\exp(iN\varphi)\ket{0}_A\ket{N}_B  \Big),
\end{equation}
where $\varphi$ is the phase per photon introduced in one of the modes (subsystems $A$ or $B$), and $N$ is the non-zero number of photons in either of the modes. N00N states can be used to estimate an unknown phase $\varphi$ with a precision that scales as $1/N$ \cite{Mitchell_2004}. Compared with protocols that make use of coherent states, that provide a precision that scales as $1/\sqrt{N}$, N00N states are an important example of quantum-enhanced phase estimation.

\begin{figure}[t!]
\includegraphics[width=\linewidth]{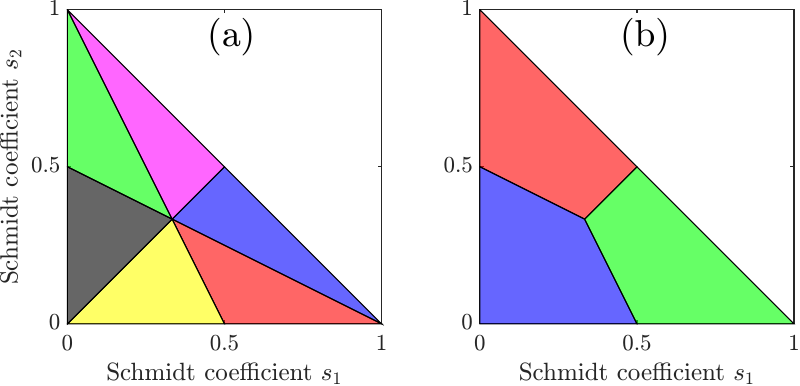}
\caption{\label{6color} \textcolor{black}{Comparison of the values of the minimum of the quantum uncertainty $U$ obtained for different Schmidt coefficients $s_1$ and $s_2$.  (a) The eigenvalues associated to the von Neumann measurement are $\lambda_1=2$, $\lambda_2 = 4$ and $\lambda_3 = 1$; (b) The eigenvalues associated to the von Neumann measurement are $\lambda_1=4$, $\lambda_2 = 3$ and $\lambda_3 = 2$. Each color designates a given value of the minimum of $U$. In (a) the value of the minimum of $U$, for all possible coefficients $s_1$ and $s_2$, can yield up to six different results. In (b) one finds only three different values of the minimum of $U$. The Schmidt coefficients are dimensionless.}}
\end{figure}

\begin{figure}[t!]
\includegraphics[width=\linewidth]{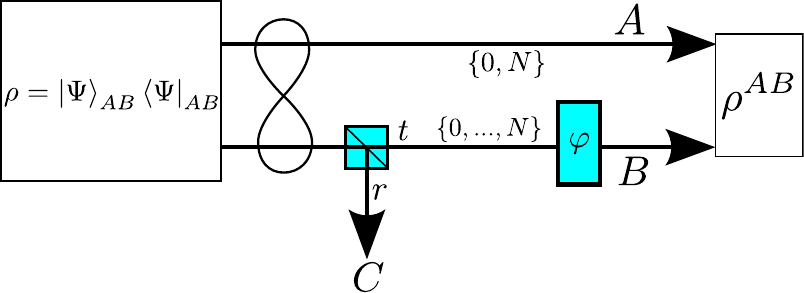}
\caption{\label{CartoonNOON} We consider the case of a simple $2\times D$ quantum system with $D=N+1$: A bipartite N00N state  $\rho^{AB}$ with non-symmetric losses. Photon losses are considered in subsystem $B$, where a fictitious beam splitter with reflectivity $r$ and transmissivity $t$ models
the system losses. Subsystem $A$ has only two accessible states $\{0,N\}$ whereas subsystem $B$ has $D=N+1$ accessible states $\{0, ... ,N\}$ given by the combination term in Eq.(\ref{ABC_combinatory}).}
\end{figure}

We consider the case where there are losses only in subsystem B (non-symmetric losses). The reason for this is that in this scenario the quantum state is a $2 \times (N+1)$ system, which allows us to calculate the quantum discord in a straightforward way. As shown in Figure \ref{CartoonNOON}, we can model such losses by considering that photons travelling in subsystem $B$ traverse a fictitious beam splitter (BS) with reflection coefficient $r$ (photons moving from subsystem $B$ to subsystem $C$) and a transmission coefficient $t$ (photons that continue in subsystem $B$) \cite{walmsleyPRA}. The overall quantum state after the BS is
\begin{eqnarray}
&  & \ket{\Psi}_{ABC}=\frac{1}{\sqrt{2}} \big[ \ket{N}_{A} \ket{0}_{B} \ket{0}_C   \label{overall} \label{ABC_combinatory} \\
&  &  +  \sum_{n=0}^N \sqrt{\binom{N}{n}}\,t^n r^{N-n} \exp (in\varphi) \ket{0}_{A} \ket{n}_{B} \ket{N-n}_C \big], \nonumber 
\end{eqnarray}
with two accessible states for subsystem $A$ ($\{0,N\}$) and $N+1$ for subsystem $B$ ($\{0, ... , N\}$).

The density matrix that describes subsystem $AB$ is obtained calculating the partial trace of the state given by Eq. (\ref{overall}) with respect to subsystem $C$. In this way,
\begin{eqnarray}
& & \rho^{AB}=\frac{1}{2}\big(\ket{N}_A \ket{0}_B+t^N \exp (iN\varphi)\ket{0}_A\ket{N}_B \Big) \nonumber \\
& & \times \Big( \bra{N}_A \bra{0}_B+t^{*N} \exp(-iN\varphi) \bra{0}_A\bra{N}_B \big) \nonumber \\ 
& & +\frac{1}{2}\sum_{n=0}^{N-1} \binom{N}{n} |t|^{2n} |r|^{2(N-n)} \ket{0}_A \ket{n}_B\bra{0}_A\bra{n}_B.
\label{partialAB}
\end{eqnarray}
The fact that the dimension of the quantum state of subsystems $AB$ is $2 \times D$ with $D=N+1$ allows us to readily calculate the Local Quantum Uncertainty, or equivalently the geometric quantum discord. 

\subsection{Calculation of the quantum Fisher information}
The quantum Fisher Information $F_Q$ associated to the quantum state given by Eq. (\ref{partialAB}) can be calculated by making use of the spectral decomposition of the state:  $\rho^{AB}=\sum_i \lambda_i(\varphi) \ket{\lambda_i(\varphi)}_{AB} \bra{\lambda_i(\varphi)}_{AB}.$ Here $\lambda_i(\varphi)$ are the eigenvalues of the decomposition and $\ket{\lambda_i(\varphi)}_{AB}$ are the corresponding eigenvectors. It can be easily demonstrated that all eigenvalues show no dependence on the value of $\varphi$ and that there are two eigenvectors with a non-zero $\varphi$ -dependence:
\begin{equation}
\left|\lambda_{1}\right\rangle ={\cal N}\left[|N\rangle_{A}|0\rangle_{B}+t^{N}\text{e}^{iN\varphi}|0\rangle_{A}|N\rangle_{B}\right],
\end{equation}
with $\lambda_{1}=\left(1+|t|^{2N}\right)/2$, and 
\begin{equation}
\left|\lambda_{2}\right\rangle ={\cal N}\left[-t^{*N}|N\rangle_{A}|0\rangle_{B}+\text{e}^{iN\varphi}|0\rangle_{A}|N\rangle_{B}\right]
\end{equation}
 with $\lambda_{2}=0$. The normalization constant is ${\cal N}=(1+|t|^{2N})^{-1/2}$. In this case \cite{walmsleyPRA,EntangledCoherentStates2013} the quantum Fisher information reads $F_Q=\lambda_1 F_1$ with
\begin{equation}
F_{1}=4\left[\bigg\langle\frac{\partial\lambda_1}{\partial\varphi}\bigg|\frac{\partial\lambda_1}{\partial\varphi}\bigg\rangle-\left|\bigg\langle\lambda_{1}\bigg|\frac{\partial\lambda_1}{\partial\varphi}\bigg\rangle\right|^{2}\right],
\end{equation}
which yields the simple expression 
\begin{equation}
    F_Q=N^2\frac{2|t|^{2N}}{1+|t|^{2N}}.
\end{equation}
Note that for the ideal lossless case, we obtain the well-known result $F_Q=N^2$.

\begin{figure}[t!]
\includegraphics[width=\linewidth]{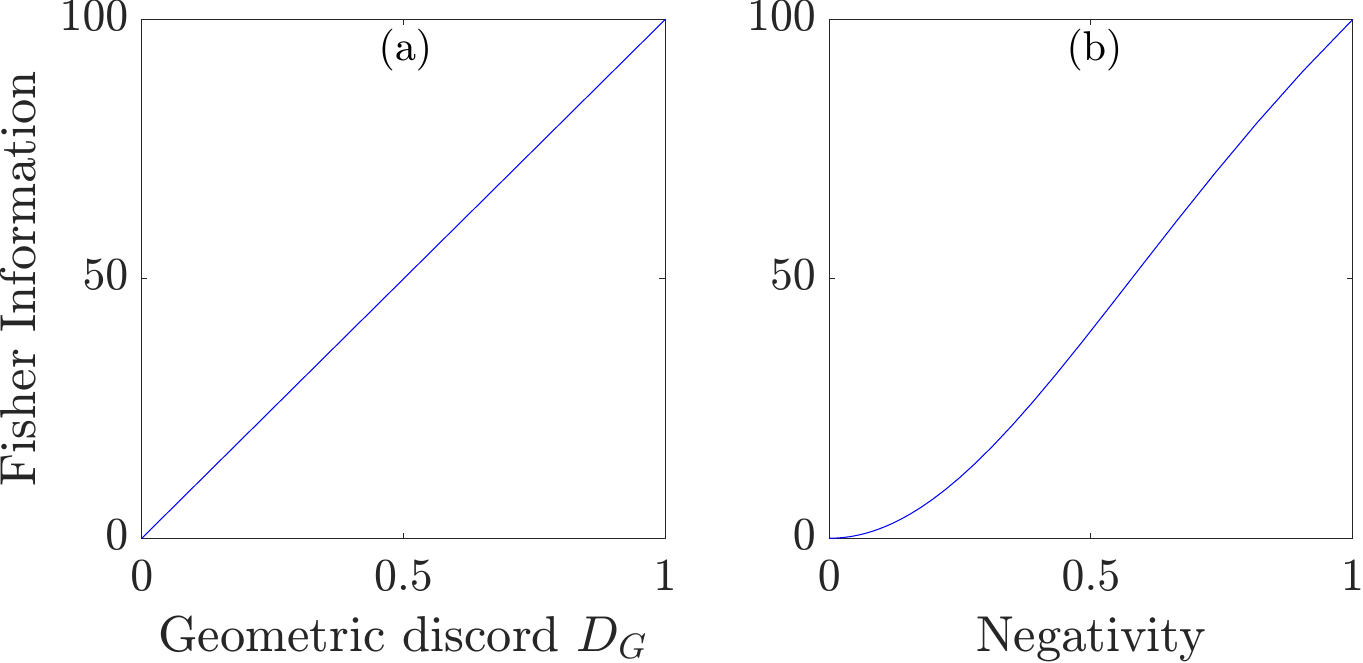}
\caption{\label{figure2} Quantum Fisher Information for the N00N state with losses in one subsystem, as a function of geometric quantum discord $D_G$ (left) and Negativity (right).}
\end{figure}

\subsection{Calculation of LQU and GQD}
Given that LQU and geometric quantum discord are equivalent discord measures for $2 \times (N+1)$ quantum systems, in what follows we will refer to them as geometric quantum discord $D_G$ for the sake of simplicity. According to Ref. \cite{Girolami2013}, the LQU of $2 \times (N+1)$  bipartite quantum systems is $D_G=1-\lambda_{\mathrm{max}}$ where $\lambda_{\mathrm{max}}$ is the greatest eigenvalue of the  $3 \times 3$ symmetric matrix $W_{AB}$,
\begin{equation}
    (W_{AB})_{ij}=\mathrm{Tr}\Big(\rho^{1/2}(\sigma_i \otimes \openone)\rho^{1/2}(\sigma_j \otimes \openone) \Big).
    \label{matrixW}
\end{equation}
Here $\sigma_i$ designates the three Pauli matrices. We obtain that the greater eigenvalue of the matrix $W$, considering the quantum state $\rho_{AB}$ described by Eq. (\ref{partialAB}), is $\lambda_{\mathrm{max}}=(1-|t|^{2N})/(1+|t|^{2N})$. Therefore the corresponding geometric quantum discord is
\begin{equation}
    D_G=\frac{2|t|^{2N}}{1+|t|^{2N}}.
\end{equation}
We can thus write a very simple relationship between the quantum Fisher information with and without loss 
\begin{equation}
    F_Q^{\text{loss}}=D_G \times F_Q^{\text{lossless}}.
\end{equation}
where $F_Q^{\text{loss}}$ designates the quantum Fisher information of the N00N state in a lossy environment and $F_Q^{\text{lossless}}$ is the quantum Fisher information of the ideal (no losses) N00N state.  Remarkably, we have found that the geometric quantum discord (and so the Local quantum uncertainty) quantifies the loss of quantum Fisher information due to losses. Fig. \ref{figure2}(a) shows the linear relationship between Fisher information and $D_G$ for a N00N state with $N=10$. It turns out that the geometric quantum discord is the decrease of quantum Fisher information of a N00N state due to non-symmetric losses.

The quantum state given by Eq. (\ref{partialAB}) is always entangled. This can be demonstrated calculating the negativity, that is an entanglement monotone \cite{horodecki2009}. Fig. \ref{figure2}(b) shows the Quantum Fisher Information as a function of negativity. For high degree of entanglement (low losses and thus negativity close to 1) the Fisher information is a quasi-linear function the negativity of the quantum state. However, for low values of entanglement (high losses and low values of negativity) the relationship between quantum Fisher information and negativity is no longer lineal, contrary to the case of the geometric quantum discord.

\section{Conclusions}
We have demonstrated that two measures of quantum discord, namely the geometric quantum discord introduced in \cite{ChangLuo2013} and the local quantum uncertainty \cite{Girolami2013} are equivalent measures of discord for $2 \times D$ quantum bipartite systems. Contrary to other measures of discord \cite{ollivier2002,vedral2011} that are very difficult to compute, these measures can be computed in closed form for $2 \times D$ systems, which include important cases such as two-qubits systems.

As an example of the relevance of the geometric quantum discord (and local quantum uncertainty), we have considered N00N states in non-symmetric lossy environments, that are $2 \times (N+1)$ quantum bipartite systems. We have found that the geometric quantum discord faithfully quantifies the decrease of quantum Fisher information due to losses, a good indicator of the quantum enhancement provided by N00N states for phase estimation.

\section*{Acknowledgements}
We acknowledge support from the Spanish Ministry of
Economy and Competitiveness (“Severo Ochoa” program for
Centres of Excellence in R\&D No. SEV-2015-0522), from Fundacio Privada Cellex, from Fundacio Mir-Puig, and from Generalitat de Catalunya through the CERCA program. This work was partially funded through the EMPIR project 17FUN01-BeCOMe. The EMPIR initiative is co-funded by the European Union Horizon 2020 research and innovation programme and the EMPIR participating States. 
A.V. thanks the financial support from PREBIST that has  received  funding  from  the  European Union’s Horizon   2020  research and innovation   programme under the Marie Sklodowska-Curie grant agreement No 754558. J.R.A. acknowledges funding by the European Union Horizon 2020 (Marie Sklodowska-Curie 765075-LIMQUET). R.J.L.M. thankfully acknowledges financial support by CONACyT under the project CB-2016-01/284372, and by DGAPA-UNAM under the project UNAM-PAPIIT IN102920.

%

\end{document}